\documentclass[]{article}

\usepackage{graphicx}
\usepackage{amsmath, amssymb}
\usepackage{booktabs}
\usepackage{subcaption}
\usepackage{abstract}
\usepackage{authblk}
\usepackage{natbib}
\begin{document}
\title{Bayesian Quantile-Based Joint Modelling of Repeated Measurement and Time-to-Event data, with an Application
to  Lung Function Decline and Time to Infection in Patients with Cystic Fibrosis}
\author[1]{Elisabeth Waldmann} 
\affil[1]{ Department of Medical Informatics, Biometry and Epidemiology\\

 Friedrich-Alexander-Universit\"at Erlangen-N\"urnberg, Germany}

\author[2]{David Taylor-Robinson} 
\affil[2]{ Department of Public Health and Policy\\

Farr Institute, University of Liverpool, UK
 }

\maketitle

\setlength{\parindent}{0pt}
 \begin{onecolabstract}
\textbf{Background} The most widely used approach to joint modelling of repeated measurement and time to event data  is to combine a linear Gaussian random effects model for the repeated measurements  with a log-Gaussian frailty model for the time-to-event outcome, linking the two through some form of correlation structure  between the random effects and the log-frailty. In this approach, covariates are assumed to affect the mean response profile of the repeated measurement data. \\
\textbf{Objectives} Some applications raise substantive questions  that cannot be captured by this structure. For example, an important question in cystic fibrosis (CF) research is to understand the impact of a patient's lung  function trajectory on their risk of acquiring a variety of infections, and how  this varies at different  quantiles of the lung function distribution. \\
\textbf{Methods} Motivated by this question, we develop a joint quantile modelling framework in this paper with an associated Markov Chain Monte Carlo algorithm for Bayesian inference.\\
\textbf{Results} The translation from the common joint model towards quantile regression succeeds and is applied to CF data from the United Kingdom. The method helps detecting an overall difference in the relation between lung function decline and onset of infection in the different quantiles.\\ 
\textbf{Conclusions} Joint modelling without taking into account the special heteroscedastic structure is not sufficient in certain research question and the extensions towards models beyond the mean is necessary.\\
\end{onecolabstract}
\section{Introduction}\label{sec:one}

Amongst the many statistical methods that have been developed for an\-lysing data from longitudinal studies, the term {\it joint modelling}  refers to the statistical modelling of data in which each subject provides data on two qualitatively different kinds of outcome variable: a time-sequence of repeated measurements; and a (possibly right-censored) time-to-event variable. The extensive literature on this topic is reviewed in \cite{tsiatis}, and in a recent text by \cite{riz_book} and \cite{asar}. Most of the joint modelling literature adopts a hierarchical modelling approach in which the repeated measurement and time-to-event outcomes are modelled as conditionally independent linear Gaussian and log-linear proportional hazards models given a latent bivariate stochastic Gaussian process, say $\{W_1(t), W_2(t)\}$ whose components affect the conditional mean of the measurement process and conditional log-hazard of the time-to-event process, respectively.\\

In  practice, this approach is usually associated with an inferential focus on how a subject's repeated measurement process affects their prognosis for survival, after adjustment for covariate effects. Quantile regression methods \citep{KoeBas1978}, in  contrast, are designed to answer questions concerning the relationship between a subject's covariates and the corresponding quantile of their measured outcomes. Papers that describe quantile regression methods for repeated measurement outcomes include \cite{koenker2004a} and \cite{Geraci} in a likelihood based scenario, \cite{Fenske} in the boosting context, and \cite{Yue} using a Bayesian approach. To our knowledge, the one paper that extends this to joint modelling as defined above is \cite{farm_viv}.\\

In this paper, we develop a novel approach to quantile-based joint modelling, motivated by a question in the epidemiology of cystic fibrosis, a genetic condition that leads to a progressive deterioration of a patient's lung function  throughout their life.\\

In Section \ref{sec:two} of the paper we give a description of cystic fibrosis, the specific research question that motivated this work and the data that we will use to answer the question. In Section
 \ref{sec:three} we set out our proposed methodology. We formulate a location scale mixture representation of the asymmetric Laplace distribution whose likelihood can be treated like that of  a latent Gaussian distribution, thus allowing the use of tools previously developed for joint modelling based on mean regression. We also develop an MCMC algorithm for Bayesian inference.   
In Section \ref{sec:four} we describe our analysis of the cystic fibrosis data. Section \ref{sec:outlook} discusses some limitations of the current methodology and outlines further work
to extend its scope. Code is available from the first author.

\section{Outline of the problem}\label{sec:two}

Cystic fibrosis (CF) is one of the most common serious genetic diseases in the Western world. It has an impact on a number of organs, primarily the lung, pancreas and liver. The disease is characterized by recurrent lung infection and inflammation with associated long-term lung function decline. Most people with cystic fibrosis die prematurely as a result of respiratory failure.

Most epidemiological studies of lung function in CF have concentrated on investigating how a range of risk factors affect the decline in mean  lung function in children and adult populations \citep{konstan2007, salvatore2011, konstan2012, salvatore2012, dtr2012, dtr2013}. A number of studies have found that infection with \textit{Pseudomonas aeruginosa} (PA) has an accelerating effect on the
 decline in mean lung function \citep{rosenfeld, dtr2012}. However, this acceleration is potentially greater and thus more concerning for patients 
whose  lung function is already relatively poor, i.e. for individuals  whose lung function is at lower quantiles of the distribution. A first step to further explore the impact of PA on lung function is to set the problem in a quantile regression context \citep{KoeBas1978}. Instead of assessing the impact of a set of covariates on the mean, we analyze their impact on selected quantiles. This provides additional insights  without the need to assume a closed form distribution; we give a more detailed description in Section \ref{sec:three}.

Quantile regression has been used rarely in research on pulmonary diseases. Examples include \cite{sickle}, who showed differences in lung function decline for groups with different socioeconomic background, \cite{kameryn}, who modelled lung function decline in CF, and \cite{kulich}, who estimated reference equations to be able to classify the lung function level of CF patients in comparison to other CF patients, rather than to a healthy population.

In this paper we use data from the UK CF registry, which collects longitudinal data on all patients with CF living in the UK. The registry currently collects data on around 10,000 patients who attend annual examinations examinations where they are assessed for clinical status, pulmonary function and microbiology of lower respiratory tract secretions. Lung function is measured using forced expiratory volume in 1 second as a percentage of predicted (\%FEV$_1$) which is an age-standardised measure of lung function. We analyse an extract from the registry, containing a total of 16,872 \%FEV$_1$ measures from the 3,200  patients who were seen at least twice between 1995 and 2009. Of the 3,200 patients, 1,237 (39\%) were infected with PA at some point during the follow-up period. We aim to develop a better understanding of the connection between the different levels of decline in the lung function and the risk of PA acquisition. 

\begin{figure}[h!]\centering
\includegraphics[width=\textwidth]{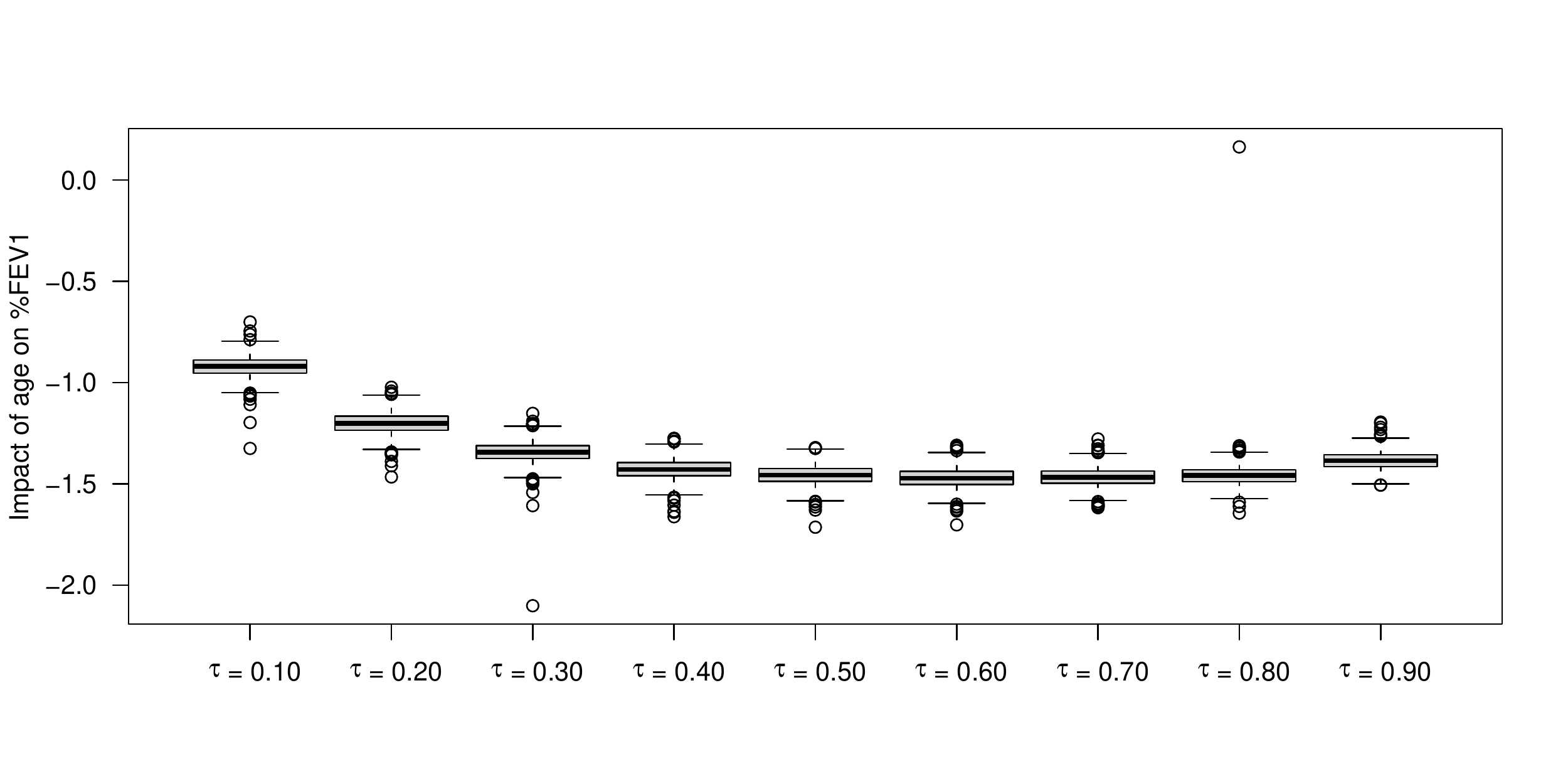}
\caption{\label{box_quant} MCMC samples from the posterior distribution of  $\beta_{\text{PA}}$ in quantile regression at nine different quantiles, $\tau$}
\end{figure}

A quantile regression analysis of the repeated measurements of lung function reveals clear differences in the impact of PA at different quantiles of \%FEV$_1$. We fitted the following model,
\begin{equation}
\boldsymbol{y}_{\tau} = \beta_0 + \boldsymbol{x}_{\text{age}}\beta_{\text{age}} + \boldsymbol{x}_{\text{PA}}\beta_{\text{PA}} + \boldsymbol{\gamma}_0 + \boldsymbol{\gamma}_{\text{age}}\boldsymbol{x}_{\text{age}}, 
\label{simp_quant}
\end{equation}
where, for each patient,  $\boldsymbol{x}_{\text{age}}$ is the vector of the age at which \%FEV$_1$ was measured, $\boldsymbol{x}_{\text{PA}}$  contains zeros until the onset of infection and years of  infection thereafter, $\boldsymbol{\gamma}_0$ and $\boldsymbol{\gamma}_{\text{age}}$ are random intercept and slope terms, and $\boldsymbol{y}_{\tau}$ is the $\tau$-quantile of \%FEV$_1$. For inference on this model, we used the Bayesian software tool BayesX (\cite{bayesx}).
Figure ~\ref{box_quant} shows that the impact of PA on lung function decline, as measured by the parameter $\beta_{\text{PA}}$,
 is less pronounced at central and higher quantiles of the distribution. This suggests that infection with PA has a bigger impact on lung function for patients with worse preexisting lung function and constitutes a prima facie case for investigating  structural differences in the relationship between lung function and PA infection at different quantiles of \%FEV$_1$.

Joint modelling of the kind described in Section 1  has previously been used to investigate  cystic fibrosis data for the relationship between lung function decline and survival of CF patients \citep{barrett, schluchter}. To best of our knowledge, the present paper is the first attempt to use quantile regression methods to understand the relationship between lung function decline 
and risk of PA infection.

\section{ Quantile-based joint modelling }\label{sec:three}

\subsection{Mean Regression Bayesian Joint Modelling}\label{BJM}

We first set out an example of a mean regression joint model proposed in
 \cite{faucett}, which provides the starting point  for our qunatile-based
 methodology and our associated  Monte Carlo Markov Chain (MCMC) 
 algorithm. The model specification is
\begin{eqnarray}
y_{ij}&=& \eta_{ij, ls}(t_{ij}, x_{ij}) + \eta_{ij, l}(t_{ij}, x_{ij}) + \varepsilon_{ij}: j=1,...,n_i; i=1,...,n\nonumber\\
\varepsilon_{ij}&\overset{\text{iid}}{\sim}&N(0,\sigma^2)\label{JM}\\
\hat{\lambda}_i(t_{ij})& =& \lambda_{0}(t_{ij})\exp(\alpha\eta_{ij, ls}(t_{ij}, x_{ij}) + \eta_{ij, s}(t_{ij}, x_{ij})),\nonumber\\\nonumber
\end{eqnarray}
where $y_{ij}$ is the $j$th repeated measurement 
on the $i$th of $n$ individuals,  $t_{ij}$ is the corresponding measurement time and
the $\eta_{ij, \cdot}(t_{ij}, x_{ij})$ are specified
functions of $t_{ij}$, covariates $x_{ij}$ and  individual-level random effects that
have an impact on the repeated measurements (subscript $l$), the hazard function (subscript $s$)
or both (subscript $ls$).

\begin{eqnarray*}
\eta_{ij, s}(t_{ij}, x_{ij})&=& 0\\\nonumber
 \eta_{ij, l}(t_{ij}, x_{ij})&=& \beta_{0} + \beta_{1} t_{ij}\\\nonumber
\eta_{ij, ls}(t_{ij}, x_{ij})&=& \gamma_{0i} + \gamma_{1i} t_{ij}\\\nonumber
\end{eqnarray*}
To estimate the distribution of the survival times we use the partial likelihood for the hazard rate. Since the composition of the predictor is rather messy we present the parts in a structured form.

\begin{description}
\item[$\eta_{ij, ls}(t_{ij}, x_{ij}) = $]  $\sum_{k=1}^{p} f_{k, ls}(t_{ij}, x_{ij}, \boldsymbol{\theta}_{ls})$ where the $f_{k, ls}(t_{ij}, x_{ij}, \boldsymbol{\theta})$ in the sum can be different kinds of functions of the covariates $x_{ij}$ and the time $t_{ij}$. The parameter vector $\boldsymbol{\theta}_{ls}$ includes all parameters necessary for the corresponding functions. The index $ls$ indicates that this part of the predictor is assumed to be influential for both parts of the model, the {\textit {longitudinal outcomes}} and the {\textit {survival times}}. In the survival part of the model this part of the predictor is multiplied by the association parameter $\alpha$.
\item[$\eta_{ij, l}(t_{ij}, x_{ij}) = $]  $\sum_{k=1}^{p} f_{k, l}(t_{ij}, x_{ij}, \boldsymbol{\theta}_{l})$. The structure of this function is the same as above, but it contains the covariates being assumed to only be relevant for the {\textit{longitudinal part}} of the model.
\item[$\eta_{ij, s}(t_{ij}, x_{ij}) = $]  $\sum_{k=1}^{p} f_{k, s}(t_{ij}, x_{ij}, \boldsymbol{\theta}_{s})$. The structure of this function is the same as above, but it contains the covariates being assumed to only be relevant for the {\textit{survival part}} of the model.
\end{description}
To achieve better legibility we present the functions without the corresponding arguments (i.e. $\eta_{ij, ls}$ instead of $\eta_{ij, ls}(t_{ij}, x_{ij})$), unless the arguments differ from the $t_{ij}$ and $x_{ij}$. This is the case e.g. in the second line of the following formula, where the function is evaluated at the survival times or the third line, where the argument of the formula is the integration variable. 
The likelihood for the model of structure (\ref{JM}) results in:
\begin{eqnarray}
f(Y_{ij}|\eta_{ij, ls}, \eta_{ij, l},\sigma^2) &=&\frac{1}{\sqrt{2\pi\sigma^2}^{J_i}}\exp\left(-\sum_{j=1}^{J_i}\frac{\left(y_{ij}-(\eta_{ij, ls}+\eta_{ij, l})\right)^2}{\sigma^2}\right)\nonumber\\
f(s_i,d_i|\eta_{ij, ls}, \eta_{ij, s}, \alpha)&=&\left\{\lambda_0(s_i)\exp\left(\eta_{ij, ls}(s_i, x_{ij})+ \eta_{ij, s}(s_i, x_{ij})\right)\right\}^{d_i}\label{JM_LI}\\
&&\exp\left(\int_{e_i}^{s_i}\lambda_0(u)\exp\left(\alpha\eta_{ij, ls}(u, x_{ij})+\eta_{ij, s}(u, x_{ij})\right)du\right),\nonumber\\\nonumber
\end{eqnarray}
where $n$ is the number of individuals, $J_i$ the number of observations of the $i$-th individual, $e_i$ is the time individual $i$ enters the study and $s_i$ is either the time of event or last time observation. The dependent variable $y_{ij}$ is the $j$-th observation of individual $i$. To cope with the baseline hazard $\lambda_0(t)$, the time is split into a grid $t_0,\cdots, t_K$ for which $\lambda_0(t)$ is approximated by a piecewise constant function with values $\lambda_k$ for $k = 1,\cdots,K$. The integral in equation (\ref{JM_LI}) can hence be approximated by a sum of $K$ integrals. The association parameter $\alpha$ links the mixed model to the survival analysis and quantifies the strength of the influence. In the following we will give a short overview over the full conditionals and refer to the appendix for more detailed calculation. We present the full conditionals exemplary for linear effects (e.g. $\eta_{ij, l}=\sum_{k=1}^{K}\boldsymbol{X}_k\boldsymbol{\beta}_{l,k}$), for possible extensions see section \ref{sec:outlook}.\

\paragraph{Predictor part only concerning longitudinal data:}
The predictor $\eta_{ij, l}$ is independent of the survival part of the model. When choosing a Gaussian prior for the parameters the resulting full conditional is also of Gaussian type. Just as in standard Bayesian longitudinal modelling, inference on random effects can be implemented by inducing the correlation within the individual/group by using an appropriate prior covariance matrix.

\paragraph{Predictor part only concerning survival data:}
The approach is reverse to the above presented. The likelihood for the parameters collected in $\boldsymbol{\beta}_s$ can be presented independently of the longitudinal process. The full conditional can hence be constructed as if we were only doing a Cox regression. Unfortunately there is no closed form distribution for the full conditional and we follow \cite{faucett} doing adaptive rejection sampling (ARS). 

\paragraph{Predictor part concerning both longitudinal and survival data:}
The full conditional for the predictor part $\eta_{ij, ls}$ is even more complicated than the survival specific part, since it appears in both parts of the likelihood. We hence follow \cite{faucett} again and use an ARS step.

\paragraph{Model variance:}
The model variance $\sigma^2$ is not linked to the survival part of the model, the full conditional is hence conform with the result in Bayesian mixed models.

\paragraph{Baseline Hazard:}
The relevant part of the likelihood (\ref{JM_LI}) for the full conditional of the  $\lambda_k$ is the second part, which reduces to a gamma distribution.

\paragraph{Association Parameter:}
The full conditional for $\alpha$ consists of the second term of the likelihood and is also generated by ARS. 

\subsection{Bayesian Quantile Regression}\label{qr}

Our proposed adaptation of the above framework to quantile-based joint modelling is as follows:

\begin{eqnarray}
y_{ij}&=& \eta_{ij, ls, \tau}(t_{ij}, x_{ij}) + \eta_{ij, l, \tau}(t_{ij}, x_{ij}) + \varepsilon_{ij, \tau}: j=1,...,n_i; i=1,...,n\nonumber\\
F_{\varepsilon_{ij,\tau}}(0)&=&\tau\nonumber\\
\hat{\lambda}_{i,\tau}(t_{ij})& =& \lambda_{0}(t_{ij})\exp(\alpha_{\tau}\eta_{ij, ls,\tau}(t_{ij}, x_{ij}) + \eta_{ij, s}(t_{ij}, x_{ij})),\nonumber\\\nonumber
\end{eqnarray}

where everything is just as in model \ref{JM}, except for the error not being distributed with a Gaussian distribution and the outcome not being the mean, but a quantile. The function $F_{\epsilon_{ij,\tau}}(\cdot)$ denotes the cumulative distribution function of $\varepsilon_{ij,\tau}$. We hence have one model for each quantile of interest $\tau$. Before we explain, how this is included into the joint modelling framework, we sketch out the basic idea of inference in quantile regression.
Quantile regression can be estimated by minimising the so called check function: 
\begin{eqnarray}              
\rho_{\tau}(y_{ij},\eta_{i,\tau}(x_{ij}, t_{ij}))&=&\left\{
\begin{array}{ll}
\tau(y_{ij}-\eta_{i,\tau}(x_{ij}, t_{ij}))&\text{if }y_{ij}\geq\eta_{i,\tau}(x_{ij}, t_{ij})\nonumber\\ 
(1-\tau)(y_{ij}-\eta_{i,\tau}(x_{ij}, t_{ij}))&\text{if }y_{ij}<\eta_{i,\tau}(x_{ij}, t_{ij}).\nonumber
\end{array}
\right.
\end{eqnarray}  
which puts weights on the distance of the observations to the estimated line with the value $\tau$. Optimising this  criterion however is not straightforward, given the non differentiability due to the absolute value characteristics. The Bayesian alternative most commonly used, is to use the asymmetric Laplace distribution (ALD) as an auxiliary likelihood distribution: 

$$p(y_i|\eta_{i,\tau}(x_{ij}, t_{ij}), \sigma^2, \tau) = \frac{\tau(1-\tau)}{\sigma^2}\exp\left(-\frac{y-\eta_{i,\tau}(x_{ij}, t_{ij})}{\sigma^2}\right),$$
which when maximising leads to the same point estimators as minimising the checkfunction. This detour obviously inherits the problems from the checkfunction as it still contains the non differentiable part of the latter. \cite{kozumi_kobayashi} suggested to use the location scale mixture presentation of the ALD to circumvent those problems. To this end define weights that follow a exponential distribution with rate $\frac{1}{\sigma^2}$: $\omega\sim Exp\left(\frac{1}{\sigma^2}\right)$ and $Z$ a standard Gaussian random variable $Z\sim N(0,1)$. Define the auxiliary variables $\xi=\frac{1-2\tau}{\tau(1-\tau)}$ and $\phi=\frac{2}{\tau(1-\tau)}$, then Y constructed in the following way:

$$y_{ij} = \eta_{i,\tau}(x_{ij}, t_{ij}) + \xi \omega_{ij} + \phi z_i\sqrt{\omega_{ij}\sigma^2}$$
follows an $ALD(\eta, \sigma^2, \tau)$. This leads to the possibility to construct a regression model of the form:

$$y_{ij}|\eta_{ij,\tau}(x_{ij}, t_{ij}), \tau, \omega_{ij},\sigma^2\sim N\left(\eta_{i,\tau}(x_{ij}, t_{ij}) +\xi\omega_{ij}, \sigma^2\phi \omega_{ij}\right),$$
which then renders possible to conduct MCMC in the same way as done for mean regression with an extra sampling step for the above defined weights. Here a summary of the standard approach to longitudinal modelling:

\paragraph{Fixed and Random effects} The priors for the model parameters are chosen to be Gaussian with the appropriate covariance matrices. The resulting full conditional is also Gaussian, with slight changes in the parameters compared to mean regression. Hyperpriors (e.g. to perform selection algorithm) do usually not have a different impact than in mean regression. 

\paragraph{Model Variance} If choosing an inverse gamma distribution as a prior for the model variance, the full conditional is an inverse gamma distribution, too, again, with slight parameter changes, caused by offset and weights.

\paragraph{Weights} The key difference to the mean regression MCMC approach is the extra sampling step, necessary for the weights. Resulting from the location scale mixture representation of the ALD, the weights are a priori exponentially distributed with rate $1/\sigma^2$. The full conditional is an inverse Gaussian distribution. 

For a more detailed outline see e.g.\cite{waldmann} or \cite{Rue2009}.

Given the limited number of changes, having to consider in the transition from mean to quantile regression, we considered the transition from Gaussian joint modelling to Quantile regression joint modelling.

\subsection{Bayesian Quantile Regression in Joint Modelling}
The above described data set shows strong differences in the covariates when measuring the impact on different quantiles. The structure of the presented model will hence be extended to a model describing the association between a mixed model quantile regression and the survival model. The ultimate goal of this experiment is to model the survival function based on a set of quantiles mimicking the whole distribution and thus obtaining a more complete information on the relation between conditional distribution and survival time. 

To achieve this goal we use the above described location scale mixture construction of the ALD. This makes the necessary adjustments in the likelihood (\ref{JM_LI}) where the first line has to be replaced by

$$\frac{1}{\sqrt{2\pi\sigma^2\omega_{ij}\phi}^{J_i}}\exp\left(-\sum_{j=1}^{J_i}\frac{\left(y_{ij}-(\eta_{ij} + \xi \omega_{ij})\right)^2}{2\sigma^2\omega_{ij}\phi}\right).$$

The MCMC algorithm for our quantile regression joint modelling approach is hence a combination of the results from section \ref{BJM} and section \ref{qr}. The full conditionals only change slightly in parameters. For detailed information see the appendix.

\section{Analysing the cystic fibrosis registry data}\label{sec:four}

\subsection{Model formulation}

 As described in Section \ref{sec:two}, our repeated measurement outcome is  \%FEV$_1$ and our time-to-event-outcome is the time of first infection with PA. The longitudinal predictor $\eta_{l, ij}(t_{ij})$  was chosen to contain an intercept and a linear time effect as well as random intercept and time slope. The shared predictor $\eta_{ls,ij}$ was chosen to include the random effects. The list of predictors is:

\begin{eqnarray*}
\eta_{ij, s}(t_{ij}, x_{ij})&=& 0\\\nonumber
 \eta_{ij, l, \tau}(t_{ij}, x_{ij})&=& \beta_{0} + \beta_{1} t_{ij}\\\nonumber
\eta_{ij, ls, \tau}(t_{ij}, x_{ij})&=& \gamma_{0i} + \gamma_{1i} t_{ij}\\\nonumber
\end{eqnarray*}

This is a simple model formulation and by adding the linear fixed time effect to the longitudinal sub-model, but not to the survival part, we make sure that the differences in the temporal progression captured but the magnitude of the quantiles do not interfere with $\alpha$.

The model was implemented in \cite{R}, MCMC chains consisted of a sample of $10.000$ with a burn-in period of $1000$ and a thining of nine. The sampling paths showed no inconsistencies.

\subsection{Results}

\begin{figure}[h!]\centering
\includegraphics[width=\textwidth]{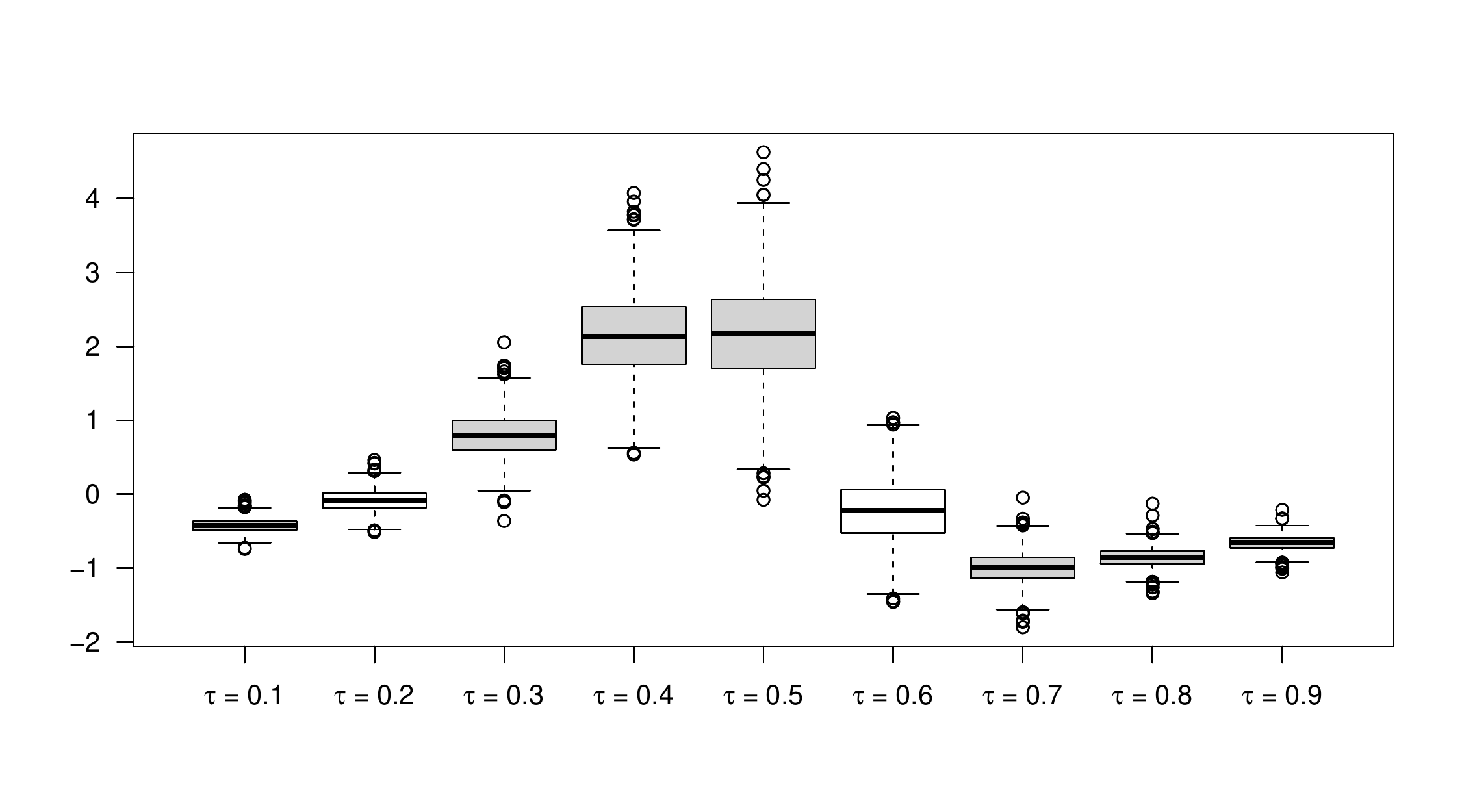}
\caption{\label{gamma} MCMC samples of the association parameter $\alpha$ for different quantiles }
\end{figure}

We ran the joint Bayesian quantile model for the nine percentiles $\tau = (0.1, 0.2, \ldots, 0.9)$. Figure~\ref{gamma} displays the different in the association parameter for the nine different quantiles. The association parameter was considered significant, if 95\% of the values in the sample are on one side of $0$ and the corresponding boxes colored grey in the figure.

\begin{figure}[h!]\centering
\includegraphics[width=\textwidth]{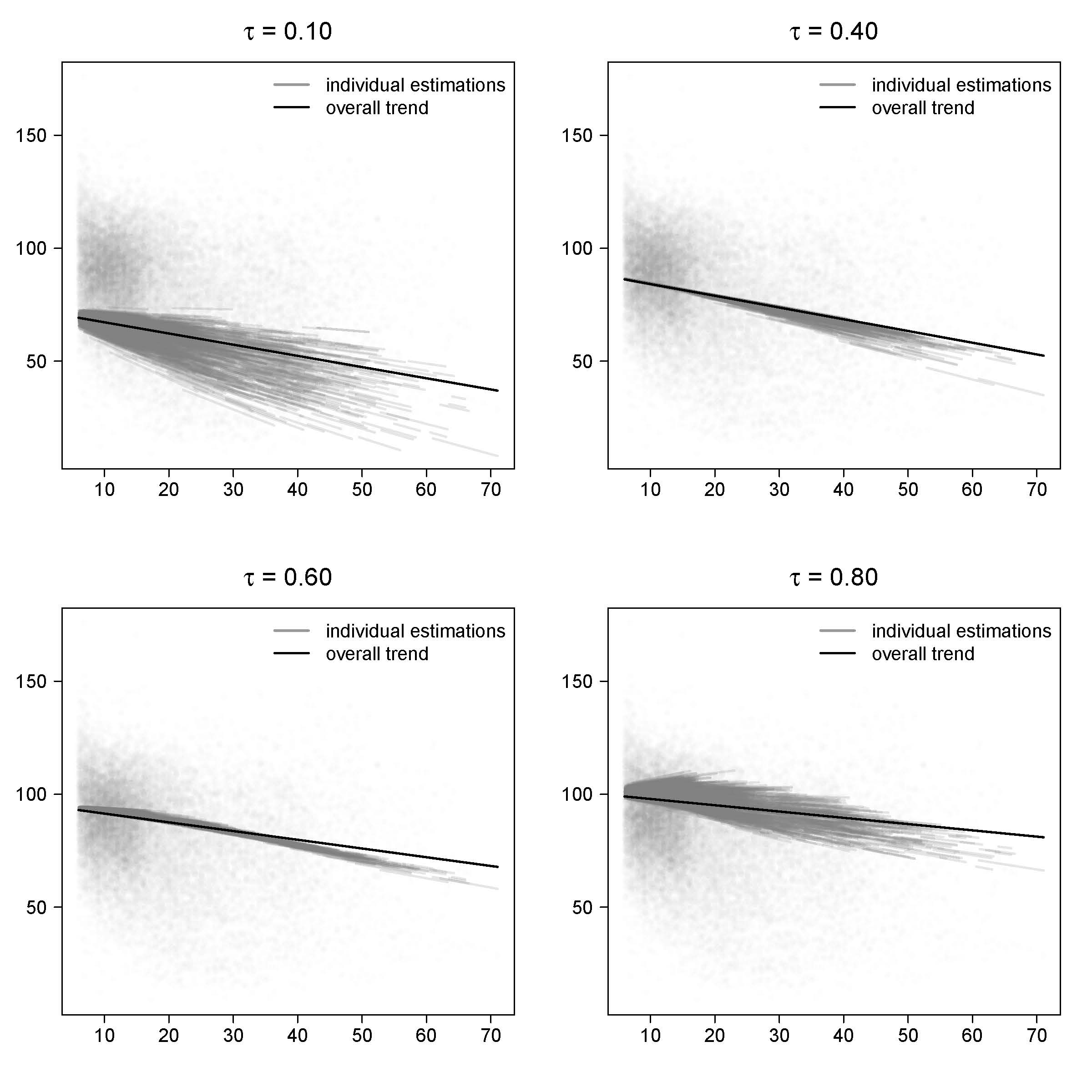}
\caption{\label{mixed} Linear effect of age on lung function with: overall effect in black, individual effects grey }
\end{figure}

To aid interpretation of these values Figure~\ref{mixed} shows a selection of the longitudinal sub-model for four different quantiles. The black lines display the overall quantile specific regression estimation, while the grey lines show the deviation from this overall estimation for individual patients. We can see that the individual lines scatter rather differently below and above the black lines. The impact of this deviation on the hazard rate is quantified by $\alpha$. Negative $\alpha$ values indicate that the patients which deviate negatively from the overall trend result in having a multiplicative constant bigger than one and hence have and increased risk of infection.  If we look at the example of $\tau=0.10$ the negative $\alpha$ (see Figure \ref{gamma}) implies that risk of infection is higher for patients, whose lines lie in the group below the dark line in the upper left figure in \ref{mixed}. Positive $\alpha$ have the opposite effect: for the more central quantiles, we can see that the $\alpha$ values indicate that the individuals above the line have a higher risk of infection whereas the risk is lower for individuals below the line. Please note for the direct interpretation that the individuals are denser around the mean trajectory for the central quantiles and the fact that the $\alpha$ values are higher does not mean that the influence is higher. Note that the individuals are not necessarily in the same group for the different quantiles. The overall conclusion is that there are differences in the relation between lung function decline and onset of infection in the different quantiles. Note that the purpose of this paper is to outline the utility of our statistical approach. Future analyses of CF data using the full dataset and taking into account a complete range of covariates will clarify the relationship between lung function and onset of infection with PA.

\section{Summary and Outlook}\label{sec:outlook}

We have shown how joint models for  repeated measurement and time-to-event outcomes  can be adapted to include ideas from quantile regression.  The proposed model could be extended in various ways. One such extension would be to use  to use lagged effects between the repeated measurement and time-to-event components of the model. A second extension would be a model in which the hazard for the time-to-event outcome depends on the rate of change in the repeated measurement process. For a random intercept and slope model, this is straightforward, we simply include the random slope as shared random effect. More generally, the model would need to include a differentiable stochastic process, $W(t)$ say, amongst the random effects in the repeated measurement component, with the derivative of $W(t)$  included in the time-to-event component.  Estimation of effects of this kind is difficult without relatively long, frequently observed series of repeated measurements on individual patients. A third extension, also straightforward in principle but difficult in practice without extensive follow-up data, would  be to allow non-linear time-varying effects.

Including more elaborate effects should be straight forward, since the location scale mixture allows for extending the predictor by numerous effects \citep{waldmann}. Besides nonlinear effect also spatial information could be included. Modelling the random effects with a Dirichlet process mixture and then including the resulting cluster information into the survival model could also lead to interesting insights into the structure of infection.

A further interesting extension would be to replace the quantile model by a Bayesian distributional regression as done by \cite{klein} or with any other kind of non mean regression in order to get a deeper understanding of the connection between a longitudinal process and time to event observations.

{\textit{Acknowledgments:} DCTR was supported by an MRC Population Health Scientist Fellowship (G0802448) for this work. We thank the UK Cystic Fibrosis Trust for access to the UK cystic fibrosis registry and all the centre directors for the input of data to the registry. EW was supported by the Interdisciplinary Center for Clinical Research (IZKF) of the Friedrich-Alexander-University Erlangen-N\"urnberg (Project J61). Special thanks go to Peter J Diggle whose comments and insights in the topic that greatly improved the manuscript.}

\clearpage

\begin{appendix}
\section{Full Conditionals}
\subsection{Bayesian Quantile Regression}

This is the likelihood setup resulting from location scale mixture explained in section~\ref{qr}:

\begin{eqnarray*}
\boldsymbol{y}|\boldsymbol{\eta},\boldsymbol{w},\tau, \sigma^2 &\sim& N(\boldsymbol{\eta}  + \xi\boldsymbol{w}, \sigma^2\phi\boldsymbol{W})\nonumber\\
w_i&\sim&Exp\left(\frac{1}{\sigma^2}\right), i=1,\ldots,n\nonumber\\
W&=&\text{diag}(w_i)\nonumber\\
\xi&=&\frac{1-2\tau}{\tau(1-\tau)}\nonumber\\
\phi&=&\frac{2}{\tau(1-\tau)}\nonumber\\
\end{eqnarray*}

We demonstrate the calculation for the example of a linear effect $\boldsymbol{\eta}_j=\boldsymbol{X}\boldsymbol{\beta}$, which forms part of the assumed additive predictor $\boldsymbol{\eta} = \sum_{j=1}^p\boldsymbol{\eta}_j$. We assign $\boldsymbol{\beta}$ a Gaussian prior $N(\boldsymbol{\mu_{\beta}},\boldsymbol{\Sigma_{\beta}})$ define the sub predictor $\boldsymbol{\eta}_{-j} = \boldsymbol{\eta}-\boldsymbol{\eta}_{j}$. Then we get a Gaussian full conditional for parameter vector $\beta$. This result is generasible towards all kinds of effects with Gaussian prior (e.g. P-Splines, Gaussian Markov random fields or random effects).

\begin{eqnarray*}
p(\boldsymbol{\beta}|\cdot)&\propto&\exp\left(\frac{1}{2}\left(\boldsymbol{\beta}-\boldsymbol{\mu_{\beta}}\right)^{\top}\boldsymbol{\Sigma}_{\beta}^{-1}\left(\boldsymbol{\beta}-\boldsymbol{\mu_{\beta}}\right)\right)\\
&&\exp\left(\frac{1}{2\sigma^2\phi}\left(\boldsymbol{X}\boldsymbol{\beta}-(\boldsymbol{y}-\boldsymbol{\eta}_{-j})\right)^{\top}\boldsymbol{W}^{-1}
\left(\boldsymbol{X}\boldsymbol{\beta}-(\boldsymbol{y}-\boldsymbol{\eta}_{-j})\right)
\right)\\
\boldsymbol{\beta}|\cdot&\sim&N(\boldsymbol{\mu}^*_{\beta}, \boldsymbol{\Sigma}^*_{\beta})\\
\boldsymbol{\mu}^*_{\beta}&=&\boldsymbol{\Sigma}^*_{\beta}\left(\frac{1}{\sigma^2\phi}\boldsymbol{W}^{-1}\boldsymbol{X}^{\top}(\boldsymbol{y}-\boldsymbol{\eta}_{-j} )+\boldsymbol{\Sigma_{\beta}}^{-1}\boldsymbol{\mu_{\beta}}\right)\\
\boldsymbol{\Sigma}^*_{\beta}&=&\left(\frac{1}{\sigma^2\phi}\boldsymbol{X}^{\top}\boldsymbol{W}^{-1}\boldsymbol{X} +\boldsymbol{\Sigma}^{-1}_{\beta}\right)^{-1}.\\
\end{eqnarray*}

We assign an inverse gamma distribution as prior to the variance parameter $\sigma^2$: $\sigma^2\sim \text{IG}(a_0, b_0)$.

\begin{eqnarray*}
f(\sigma^2|\cdot)&\propto&f(\boldsymbol{y}|\boldsymbol{\eta},\tau, \boldsymbol{W})f(\boldsymbol{W}|\sigma^2)f(\sigma^2)\nonumber\\
&\propto&\left(\frac{1}{\sigma^2\phi}\right)^{\frac{n}{2}}|\boldsymbol{W}|^{\frac{n}{2}}\exp\left\{-\frac{1}{\sigma^2}\underset{(\star)}{\underbrace{(\boldsymbol{\eta}+\xi \boldsymbol{W}-\boldsymbol{y})^{\top}\frac{1}{2\phi}\boldsymbol{W}^{-1}(\boldsymbol{\eta}+\xi \boldsymbol{W}-\boldsymbol{y})}}\right\}
\nonumber\\
&&\left(\frac{1}{\sigma^2}\right)^n\exp\left(\frac{1}{\sigma^2}\sum_{i=1}^{n} w_i\right)\nonumber\\
&&{\sigma^2}^{-a_0-1}\exp\left(-\frac{b_0}{\sigma^2}\right)\nonumber\\
&\propto&{\sigma^2}^{-(n+\frac{n}{2}+\alpha_0)-1}\exp\left(-\frac{b_0+(\star)+\sum_{i=1}^{n}w_i}{\sigma^2}\right)\nonumber\\
\end{eqnarray*}

\begin{equation*}
 \sigma^2|\cdot \sim \text{IG}\left(a_0+\frac{3n}{2}, b_0+\frac{1}{2\phi}\sum_{i=1}^{n}w_i^{-1}(y_i-\eta_i-\xi w_i)^2+\sum_{i=1}^{n}w_i\right).
\end{equation*}

The full conditionals for the weights are calculated separately for each $w_i, i=1,\ldots,n$:
\begin{eqnarray*}
 f(w_i|\cdot)&\propto&f(y_i|\eta_i,\sigma^2, w_i, \tau)f(w_i|\sigma^2)\nonumber\\
&\propto&\frac{1}{\sqrt{w_i}}\exp\left\{-\frac{1}{2}\frac{(y_i-\eta_i-\xi w_i)^2}{\phi w_i\sigma^2}\right\}
\nonumber\\
&&\left(\frac{1}{\sigma^2}\right)\exp\left(\frac{1}{\sigma^2} w_i\right)\nonumber\\
&=&\frac{1}{\sqrt{w_i}}\exp\left\{\frac{1}{2}\frac{(y_i-\eta_i)^2-2(y_i-\eta_i)\xi w_i+w_i^2(\xi^2 +2\phi)}{\phi w_i\sigma^2}\right\}\nonumber\\
&=&\frac{1}{\sqrt{w_i}}
\exp\left\{-\underset{(\star)}{\underbrace{\frac{(y_i-\eta_i)^2}{2\sigma^2\phi}}}\cdot
\frac{1-\frac{2\xi w_i}{y_i-\eta_i} +w_i^2\frac{\xi^2+2\phi}{(y_i-\eta_i)^2}}{w_i} 
\right\}\nonumber\\
&=&\frac{1}{\sqrt{w_i}}
\exp\left\{-(\star)\cdot\frac{{(w_i^{-1})}^2-\frac{2\xi}{(y_i-\eta_i)}w_i^{-1} + \frac{\xi^2+2\phi}{(y_i-\eta_i)^2}} {w_i^{-1}}\right\}\nonumber\\
\end{eqnarray*}
With the density transformation theorem follows:
\begin{eqnarray*}
p(w_i^{-1}|\cdot)&\propto&{(w_i^{-1})}^{\frac{2}{3}}
\exp\left\{-(\star)\cdot\frac{1}{w_i^{-1}}\cdot\left(w_i^{-1}- \underset{\mu_{wi}}{\underbrace{\sqrt{\frac{\xi^2+2\phi}{(y_i-\eta_i)^2}}}}\right)^2
\right\}\nonumber\\
&=&{(w_i^{-1})}^{\frac{2}{3}}
\exp\left\{ 
-\frac{1}{2(\mu_{wi})^2}\cdot\underset{\lambda_w}{\underbrace{\frac{(\xi^2+2\phi)}{\sigma^2\phi}}}
\frac{{(w_i^{-1}-\mu_{wi})}^2}{w_i^{-1}}
\right\}\nonumber\\
&\Rightarrow&w_i^{-1}\sim \text{InvGauss}(\mu_{wi},\lambda_w).\nonumber\\
\end{eqnarray*}

\subsection{Bayesian Joint Mean Regression}

This is the likelihoodsetup from section~\ref{BJM}:
\begin{eqnarray*}
f(Y_{ij}|\eta_{ij, ls}, \eta_{ij, l},\sigma^2) &=&\frac{1}{\sqrt{2\pi\sigma^2}^{J_i}}\exp\left(-\sum_{j=1}^{J_i}\frac{\left(y_{ij}-(\eta_{ij, ls}+\eta_{ij, l})\right)^2}{\sigma^2}\right)\nonumber\\
f(s_i,d_i|\eta_{ij, ls}, \eta_{ij, s}, \alpha)&=&\left\{\lambda_0(s_i)\exp\left(\eta_{ij, ls}(s_i, x_{ij})+ \eta_{ij, s}(s_i, x_{ij})\right)\right\}^{d_i}\nonumber\\
&&\exp\left(\int_{e_i}^{s_i}\lambda_0(u)\exp\left(\alpha\eta_{ij, ls}(u, x_{ij})+\eta_{ij, s}(u, x_{ij})\right)du\right),\nonumber\\
\end{eqnarray*}

To show the full conditional for effects on the longitudinal sub predictor we again demonstrate the calculation for the example of a linear effect $\boldsymbol{\eta}_{l,j}=\boldsymbol{X}\boldsymbol{\beta}$, which forms part of the assumed additive predictor $\boldsymbol{\eta}_l = \sum_{j=1}^p\boldsymbol{\eta}_{l,j}$. We assign $\boldsymbol{\beta}$ a Gaussian prior $N(\boldsymbol{\mu_{\beta}},\boldsymbol{\Sigma_{\beta}})$ define the sub predictor $\boldsymbol{\eta}_{l,-j} = \boldsymbol{\eta}_l-\boldsymbol{\eta}_{l,j}$. Then we get a Gaussian full conditional for parameter vector $\beta$. This result is generasible towards all kinds of effects with Gaussian prior (e.g. P-Splines, Gaussian Markov random fields or random effects). Please note that $\boldsymbol{\beta}$ is merely an example and indices are thus suppressed. 

\begin{eqnarray*}
p(\boldsymbol{\beta}|\cdot)&\propto&\exp\left(\frac{1}{2}\left(\boldsymbol{\beta}-\boldsymbol{\mu_{\beta}}\right)^{\top}\boldsymbol{\Sigma}_{\beta}^{-1}\left(\boldsymbol{\beta}-\boldsymbol{\mu_{\beta}}\right)\right)\\
&&\exp\left(\frac{1}{2\sigma^2}\left(\boldsymbol{X}\boldsymbol{\beta}-(\boldsymbol{y}-\boldsymbol{\eta}_{l,-j}- \boldsymbol{\eta}_{ls})\right)^{\top}\left(\boldsymbol{X}\boldsymbol{\beta}-(\boldsymbol{y}-\boldsymbol{\eta}_{l,-j}- \boldsymbol{\eta}_{ls})\right)\right)\\
\boldsymbol{\beta}|\cdot&\sim&N(\boldsymbol{\mu}^*_{\beta}, \boldsymbol{\Sigma}^*_{\beta})\\
\boldsymbol{\mu}^*_{\beta}&=&\boldsymbol{\Sigma}^*_{\beta}\left(\frac{1}{\sigma^2}\boldsymbol{X}^{\top}(\boldsymbol{y}-\boldsymbol{\eta}_{l,-j} - \boldsymbol{\eta}_{ls})+\boldsymbol{\Sigma_{\beta}}^{-1}\boldsymbol{\mu_{\beta}}\right)\\
\boldsymbol{\Sigma}^*_{\beta}&=&\left(\frac{1}{\sigma^2}\boldsymbol{X}^{\top}\boldsymbol{X} +\boldsymbol{\Sigma}^{-1}_{\beta}\right)^{-1}.\\
\end{eqnarray*}

We assign a gamma distribution as prior to the pieces of the piecewise constant baseline hazard $\lambda_k$: $\sigma^2\sim \text{G}(a_0, b_0)$.

\begin{eqnarray*}
p(\lambda_k|\cdot)&\propto&f(\boldsymbol{y}|\boldsymbol{\eta},\tau)f(\lambda_k)\nonumber\\
&&\lambda_k^{a_0+1}\exp(-b_0\lambda_k)\nonumber\\
&&\lambda_k^{\sum_{i=1}^{n}d_{ik}}\exp\left(-\lambda_k\sum_{i=1}^{n}\int_{\max(t_k,e_i)}^{\min(t_{k+1},s_i)}\exp\left(\alpha\eta_{ij, ls}(u, x_{ij})+\eta_{ij, s}(u, x_{ij})\right)du\right)\nonumber\\
&&\lambda_k^{\sum_{i=1}^{n}d_{ik}+a_0+1}\exp\left(-\lambda_k\underset{b_{\lambda_k}}{\underbrace{\left(b_0+\sum_{i=1}^{n}\int_{\max(t_k,e_i)}^{\min(t_{k+1},s_i)}\exp\left(\alpha\eta_{ij, ls}(u, x_{ij})+\eta_{ij, s}(u, x_{ij})\right)du\right)}}\right)\nonumber\\
\lambda_k|\cdot&\sim&\text{Gamma}\left(\sum_{i=1}^{n}d_{ik}+a_0,b_{\lambda_k}\right)\nonumber\\
\end{eqnarray*}

We assign an inverse gamma distribution as prior to the variance parameter $\sigma^2$: $\sigma^2\sim \text{IG}(a_0, b_0)$.
\begin{eqnarray*}
f(\sigma^2|\cdot)&\propto&f(\boldsymbol{y}|\boldsymbol{\eta}_l, \boldsymbol{\eta}_{ls}, \sigma^2)f(\sigma^2)\nonumber\\
&\propto&\left(\frac{1}{\sigma^2}\right)^{\frac{n}{2}}\exp\left\{-\frac{1}{\sigma^2}\underset{(\star)}{\underbrace{(\boldsymbol{\eta}_l + \boldsymbol{\eta}_{ls}-\boldsymbol{y})^{\top}(\boldsymbol{\eta}+ \boldsymbol{\eta}_{ls}-\boldsymbol{y})}}\right\}
\nonumber\\
&&{\sigma^2}^{-a_0-1}\exp\left(-\frac{b_0}{\sigma^2}\right)\nonumber\\
&\propto&{\sigma^2}^{-(\frac{n}{2}+\alpha_0)-1}\exp\left(-\frac{b_0+(\star)}{\sigma^2}\right)\nonumber\\
\end{eqnarray*}

\begin{equation*}
 \delta^2|\cdot \sim \text{IG}\left(a_0+\frac{n}{2}, b_0+\frac{1}{2}\sum_{i=1}^{n}(y_i - \eta_{li}-\eta_{lsi})^2\right).
\end{equation*}

\subsection{Bayesian Joint Quantile Regression}
Everything is analogous to the above. The results gets either obvious, when extending what is described in A.1 by the extra sub-predictor $\boldsymbol{\eta}_{ls}$ or when replacing the Gaussian distribution in A.2 by the location scale mixture. 
\end{appendix}

 \bibliographystyle{plainnat}
\bibliography{library}

\end{document}